\documentclass{ws-mpla}
\usepackage{graphics,axodraw,setspace,amssymb}

\begin{document}

\newcommand{\be}{\begin{equation}}
\newcommand{\ee}{\end{equation}}

\newcommand{\bea}{\begin{eqnarray}}
\newcommand{\eea}{\end{eqnarray}}

\newcommand{\ol}{\overline}

\markboth{J. K. Parry}
{{$\Delta M_s$ in the MSSM with large $\tan\beta$}}

\catchline{}{}{}{}{}

\title{$B_s^0$-$\bar{B}_s^0$ MIXING IN THE MSSM WITH LARGE $\tan\beta$}

\author{\footnotesize J. K. Parry}

\address{
Center for High Energy Physics and Department of Physics,\\ 
Tsinghua University,
Beijing 100084, P.R. of China.\\
jkparry@tsinghua.edu.cn}

\maketitle

\pub{Received 11 October 2006}{}

\begin{abstract}
The $B_s$-$\bar{B}_s$ mixing parameter $\Delta M_s$ is studied 
in the MSSM with large $\tan\beta$. 
The recent Tevatron measurement of $\Delta M_s$ is used to constrain
the MSSM parameter space.
From this analysis 
the often neglected contribution to $\Delta M_s$ from the operator
$Q^{SLL}_1$ is found to be significant.

\end{abstract}

\ccode{PACS Nos.: 14.40.Nd, 12.60.Jv}

\section{Introduction}

In the Standard Model(SM) flavour changing neutral current(FCNC) 
processes are absent at tree-level and only enter at higher orders.
In extensions of the SM there exist numerous additional 
sources of FCNC. A clear example comes from the mixings present 
in the squark sector of the Minimal Supersymmetric 
Standard Model(MSSM). These mixings will also contribute
to FCNCs at the one-loop level and could even be larger than their 
SM counterparts. An example that we shall study in this work
is the flavour changing couplings of neutral Higgs bosons 
and the neutral Higgs penguin contribution to such decays
as $B_s^0 \to \mu^+\mu^-$ and $B_s^0\!-\!\bar{B}_s^0$ mixing.
It is clear that such FCNC processes are an ideal place to 
search for physics beyond the Standard Model. Since the recent
measurement of $\Delta M_s$ at the Tevatron\cite{cdf:DMs,Abazov:2006dm}
its consequences have been studied
model independently\cite{Ligeti:2006pm,Ball:2006xx,Grossman:2006ce},
in the MSSM\cite{Ciuchini:2006dx,Endo:2006dm,Foster:2006ze,Baek:2006fq,Arnowitt:2006wn}, with minimal flavour violation\cite{MFV},
in GUTs\cite{Parry:2006mv,Dutta:2006gq}, 
in $Z^\prime$ models\cite{Cheung:2006tm,He:2006bk,Baek:2006bv},
with R-parity violation\cite{Nandi:2006qe,Xiangdong:2006sz,Wang:2006xh}, two Higgs doublet models\cite{Lu:2006xk}
and warped extra dimensions\cite{ExDim}.

In this work $B_s$-$\bar{B}_s$ mixing is studied via two methods. 
The first analysis is based on the simple SUSY SU(5) model
studied recently\cite{Parry:2005fp}. 
The second case is
that of the MSSM Higgs sector making use of the FeynHiggs numerical
package.

\subsection{$\Delta M_s$ in the large $\tan \beta$ limit}


It has been pointed out that Higgs mediated FCNC processes
could be among the first signals of supersymmetry(SUSY)\cite{Huang:1998vb,Choudhury:1998ze,Babu:1999hn}. 
In the MSSM radiatively induced couplings between the up Higgs, 
$H_{u}$, and down-type quarks may result in flavour 
changing Higgs couplings. 
In turn this will lead to large FCNC
decay rates for such process as
$B_s\to\mu^{+}\mu^{-}$ and $B_s^0\!-\!\bar{B}_s^0$ mixing. 

\begin{figure}[ht]
\begin{center}
\vskip-5mm
\rotatebox{90}{\scalebox{0.3}{\includegraphics*{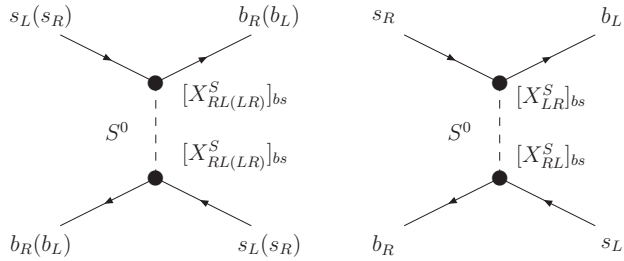}}}
\vskip-25mm
\caption{Higgs Penguin contributions to $\Delta M_s$ via the operators
$Q_1^{SLL}(Q_1^{SRR})$ and $Q_2^{LR}$.\label{fig:DP}}
\end{center}
\end{figure}
In the MSSM, loop diagrams will induce flavour changing couplings of the 
form, $b^{c}sH_{u}^{0*}$. 
Similar diagrams with Higgs fields replaced by their VEVs 
will also provide down quark mass corrections
and will lead to sizeable 
corrections to the mass eigenvalues\cite{Hempfling,Hall,Carena} and mixing 
matrices\cite{Blazek:1995nv}. 
As a result the 3-point 
coupling and mass matrix 
can not be simultaneously diagonalised\cite{Hamzaoui:1998nu}
and hence beyond tree-level we shall have flavour changing Higgs
couplings in the mass eigenstate basis. Such flavour changing Higgs
couplings can be summarized as,
\be
    {\mathcal L}_{FCNC} \;=\; 
                            -\ol{d}_{R\,i}
                             \left[X^{S}_{RL}\right]_{ij}
                             d_{L\,j} \, S^0
                            -\ol{d}_{L\,i} 
                             \left[X^{S}_{LR}\right]_{ij}
                             d_{R\,j} \, S^0\,  .
\label{L_FCNC}
\ee
These flavour changing couplings can 
in fact be related in a simple way to the
finite non-logarithmic mass matrix corrections\cite{Blazek:2003hv},
\be
\left[X^{S}_{RL}\right]_{ij} = 
\frac{1}{\sqrt{2}}\frac{1}{c_\beta}
\left(\frac{\delta m_d^{finite}}{v_u}\right)_{ij}\, A_{S^0}
\ee
where, 
$A_{S^0}=\left( s_{\alpha -\beta},\, c_{\alpha - \beta},\,-i   \right)$,
for $S^0= \left( H^0,\,h^0,\,A^0 \right)$.
It is clear that the FCNC couplings are related as, 
$\left[X_{RL}\right]=\left[X_{LR}\right]^{\dagger}$.
In general we should also notice that, 
$\left[X_{RL}\right]_{ij}\approx \frac{m_i}{m_j}\left[X_{RL}\right]_{ji}$.
Hence, in the case of, $(i,\,j)=(b,\,s)$, we have 
$\left[X_{RL}\right]_{bs}\approx \frac{m_b}{m_s}\left[X_{LR}\right]_{bs}$.

In the MSSM with large $\tan\beta$ the dominant contribution 
to $B_s\to\ell^+\ell^-$ comes from the penguin diagram where 
the dilepton pair is produced from a virtual Higgs state.
The Higgs Double Penguin(DP) contribution
to $B_s^0\!-\! \bar{B}_s^0$ mixing, shown in fig.~\ref{fig:DP} 
is also the dominant
SUSY contribution in the 
large $\tan\beta$ limit\cite{Buras:2002vd,Buras:2002wq}. 
Following the notation of eq.~(\ref{L_FCNC}),
we can write the neutral Higgs contribution to the 
$\Delta B=\Delta S=2$ effective Hamiltonian as,
\bea
{\mathcal H}_{\sf eff}^{\Delta B=\Delta S=2}
&=& 
\frac{1}{2}\sum_{S}\,\frac{[X^S_{RL}]_{bs}[X^S_{RL}]_{bs}}{-M_S^2}\,\,Q_1^{SLL}
+
\frac{1}{2}\sum_{S}\,\frac{[X^S_{LR}]_{bs}[X^S_{LR}]_{bs}}{-M_S^2}\,\,Q_1^{SRR}
\nonumber\\
&&\,\,\,\,
+\sum_{S}\, \frac{[X^S_{RL}]_{bs}[X^S_{LR}]_{bs}}{-M_S^2}\,\,Q_2^{LR}
\label{Heff}
\eea
where we have defined the operators,
\bea
Q_1^{SLL} &=& (\ol{b}P_L s)\,(\ol{b}P_L s)\nonumber\\
Q_1^{SRR} &=& (\ol{b}P_R s)\,(\ol{b}P_R s)\label{ops}\\
Q_2^{LR}  &=& (\ol{b}P_L s)\,(\ol{b}P_R s)\nonumber
\eea
The Higgs sum in eq.~(\ref{Heff}) leads to a factor, 
${\mathcal F}^{\pm}=\left( \frac{s_{\alpha-\beta}^2}{M_H^2}+
\frac{c_{\alpha-\beta}^2}{M_h^2}\pm
\frac{1}{M_A^2}  \right)$. The operators $Q_1^{SLL,SRR}$ receive
the factor ${\mathcal F}^{-}$, while $Q_2^{LR}$ receives ${\mathcal F}^{+}$.
The additional minus sign leads to a suppression of 
the $Q_1^{SLL,SRR}$ operators relative to $Q_2^{LR}$. 
At this point it is common to assume that the $Q_1^{SLL,SRR}$
contributions are negligible.
Recalling that,
$[X_{LR}]_{bs}\sim \frac{1}{40}[X_{RL}]_{bs}$, even for a suppression of
${\mathcal F}^{-}/{\mathcal F}^{+}\sim 1/100$, it may be possible
for the $Q_1^{SLL}$ contribution to give a significant effect. 
On the other hand, the contribution to $Q_1^{SRR}$ is highly suppressed.

\begin{figure}[ht]
\begin{center}
\vskip-9mm
\scalebox{0.3}{\includegraphics*{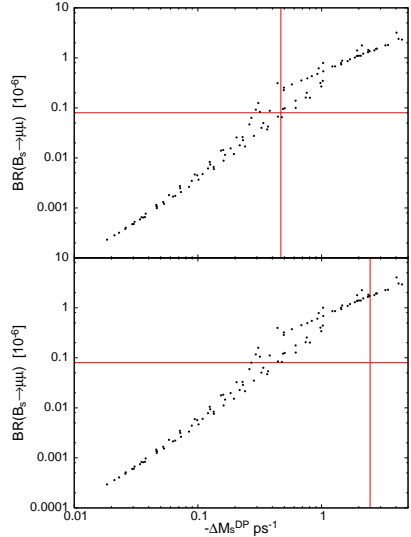}}
\vskip-11mm
\caption{
The correlation of Br$(B_s\to\mu^+\mu^-)$ 
and $\Delta M_s^{DP}$ using $f_{B_s}=230$ MeV(upper panel) 
and $f_{B_s}=259$ MeV(lower panel). 
The horizontal and vertical lines show the present 
90\% C.L. upper
bound\protect\cite{cdf:Bsmm} Br$(B_s\to\mu^+\mu^-)< 0.8\times 10^{-7}$ and 
the central value of the difference $(\Delta M_s^{CDF}\!\!-\!\Delta M_s^{SM})$
repectively.
\label{plot_DMs}
}
\end{center}
\end{figure}
Following the above conventions 
we can write the double penguin contribution to $\Delta M_s$ as,
\bea
\Delta M_s^{DP}&\equiv& 2 {Re}
\langle {\mathcal H}_{\sf eff}^{\Delta B=\Delta S=2}\rangle=
\Delta M_s^{LL}+\Delta M_s^{LR}\nonumber\\
&=&
-\frac{1}{3}M_{B_s}f_{B_s}^2 P_1^{SLL}\, 
\sum_{S}\,
\frac{[X^S_{RL}]_{bs}[X^S_{RL}]_{bs}+[X^S_{LR}]_{bs}[X^S_{LR}]_{bs}}{M_S^2}
\label{dmsDP}\\
&&-\frac{2}{3}M_{B_s}f_{B_s}^2 P_2^{LR}\, 
\sum_{S}\,
\frac{[X^S_{RL}]_{bs}[X^S_{LR}]_{bs}}{M_S^2}
\nonumber
\eea
In eq.~(\ref{dmsDP}) we have defined $\Delta M_s^{LL}$ as the contribution
from $Q^{SLL,SRR}_1$ and $\Delta M_s^{LR}$ from $Q^{LR}_2$.
Here $P_1^{SLL}=-1.06$ and $P_2^{LR}=2.56$, 
include NLO QCD renormalisation group factors\cite{Buras:2001ra}
and arise from the matrix elements of the operators of eq.~(\ref{ops}).
After taking into account the relative values of ${\mathcal F^{\pm}}$, 
the two $P$'s and the factor
of 2 in eq.~(\ref{dmsDP}), we can see that there is a relative suppression, 
\be
\frac{\Delta M_s^{LL}}{\Delta M_s^{LR}}
\approx \frac{\mathcal F^-}{\mathcal F^+}\cdot
\frac{m_b}{m_s}\cdot\frac{P^{SLL}_1}{P^{LR}_2}\cdot\frac{1}{2}
\label{suppress}
\ee
This relative suppression shall be analysed further in the following
section where it is shown that the contribution $\Delta M_s^{LL}$ is 
in fact significant.

\vskip-3mm
\begin{figure}[ht]
\begin{center}
\rotatebox{-90}{\scalebox{0.25}{\includegraphics*{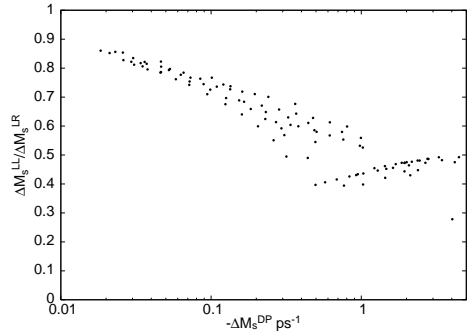}}}
\caption{\small{Plot of the ratio, $\Delta M_s^{LL}/\Delta M_s^{LR}$,
of the Higgs contributions from $Q_1^{SLL}$ and $Q_2^{LR}$.
}\label{plot_DMsLLRL}}
\end{center}
\end{figure}
There is a large non-perturbative uncertainty in the 
determination of $f_{B_s}$. Two recent lattice determinations provide
\cite{Hashimoto:2004hn,Gray:2005ad},
\bea
f_{B_s}^{'04}&=&230\pm 30 \,{\rm MeV}\label{fBs1}\\
f_{B_s}^{'05}&=&259\pm 32 \,{\rm MeV},\label{fBs2}
\eea
which in turn give different direct Standard Model predictions for 
$\Delta {M_s^{SM}}$,
\bea
\Delta M_s^{SM'04}&=&17.8\pm 8 \,{\rm ps}^{-1}\\
\Delta M_s^{SM'05}&=&19.8\pm 5.5 \,{\rm ps}^{-1}
\label{dmsSM}
\eea
The recent precise Tevatron measurement of $\Delta M_s$ 
is consistent with
these direct SM prediction but with a lower central value 
\cite{cdf:DMs,Abazov:2006dm} ,
\be
\Delta M_s^{CDF}=17.31^{+0.33}_{-0.18}\pm 0.07 \,{\rm ps}^{-1}
\label{dmsCDF}
\ee

\section{Discussion}\label{results}

We shall now discuss the Higgs mediated 
contribution to $B_s$-$\bar{B}_s$ mixing, firstly 
in a simple SU(5) SUSY GUT and secondly for the MSSM Higgs
sector using the FeynHiggs numerical package.

\vskip-3mm
\begin{figure}[ht]
\begin{center}
\rotatebox{-90}{\scalebox{0.25}{\includegraphics*{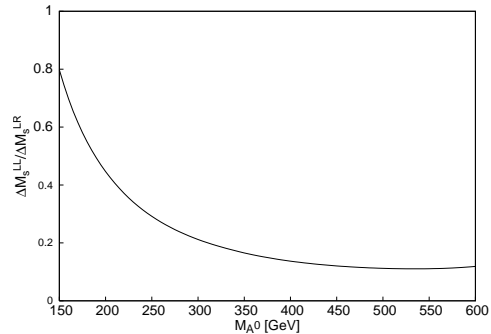}}}
\caption{A plot of the relative suppression of $\Delta M_s^{LL}$ to 
$\Delta M_s^{LR}$ against the pseudoscalar Higgs
mass. The plot is generated using the FeynHiggs package with the
input values, $\tan\beta=50$, $\mu=1000$ GeV, $M_{SUSY}=500$ GeV
and $X_t=1000$ GeV.\label{plot_var_MA0}}
\end{center}
\end{figure}
Recently a simple SUSY SU(5) model was studied using 
a top-down global $\chi^2$ analysis\cite{Parry:2005fp}.
In this model the large $\tan \beta$ 
MSSM+$N_R$ is constrained at the GUT scale by 
SU(5) unification and universal soft SUSY breaking terms.
In this work the best fits in the $(m_0,\,M_{1/2})$ parameter space
are used to make predictions for both $B_s\to\mu\mu$ and $\Delta M_s$.

In the limit of large $\tan\beta$,
$B_s\to\mu^+\mu^-$ and $\Delta M_s$ are
correlated. This correlation is shown in the two panels
of fig.~\ref{plot_DMs}. For these two panels the two different
values of $f_{B_s}$ listed in eq.~(\ref{fBs1},\ref{fBs2}) 
are used. The upper panel
($f_{B_s}=230$ MeV) shows that the central value of the difference
$(\Delta M_s^{CDF}\!\!-\!\Delta M_s^{SM})$ coincides
with the bound from Br$(B_s\to\mu^+\mu^-)$.
The lower panel ($f_{B_s}=259$ MeV) shows that the data points
with $\Delta M_s^{DP}$ at the central value, are in fact
ruled out by the bound on Br$(B_s\to\mu^+\mu^-)$. The uncertainty 
in the SM prediction
for $\Delta M_s$ is rather large and in fact all of the data points
of fig.~\ref{plot_DMs} are allowed 
by the recent Tevatron measurement at the $1\sigma$ level.
These two panels clearly show that the interpretation of the 
recent measurement depends crucially on the uncertainty in the 
determination of $f_{B_s}$. 

The plot in  fig.~\ref{plot_DMsLLRL} shows the ratio, 
$\Delta M_s^{LL}/\Delta M_s^{LR}$, 
of the contributions to 
the operators $Q_1^{SLL}$ and $Q_2^{LR}$
as defined in eq.~(\ref{dmsDP}). 
It is commonly assumed that the contribution to the $Q_1^{SLL}$
operator, $\Delta M_s^{LL}$, is negligible. From fig.~\ref{plot_DMsLLRL}
we can see that $\Delta M_s^{LL}$ is 
between 40\% and 90\% of $\Delta M_s^{LR}$ and hence is significant.

\vskip-3mm
\begin{figure}[ht]
\begin{center}
\rotatebox{-90}{\scalebox{0.25}{\includegraphics*{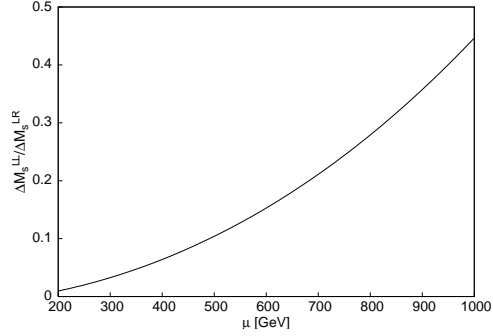}}}
\caption{A plot of the relative suppression of $\Delta M_s^{LL}$ to 
$\Delta M_s^{LR}$ against $\mu$. 
The plot is generated using the FeynHiggs package with the
input values, $\tan\beta=50$, $M_{A^0}=200$ GeV, $M_{SUSY}=500$ GeV
and $X_t=1000$ GeV.\label{plot_var_mu}}
\end{center}
\end{figure}
FeynHiggs\cite{FeynHiggs,Heinemeyer:1998yj} 
is a numerical package for computing 
the MSSM Higgs boson masses and related observables, including higher-order
corrections. Making use of this numerical package the relative
suppression of $\Delta M_s^{LL}/\Delta M_s^{LR}$ for $\tan\beta=50$
was also studied. Using the FeynHiggs package the 2-loop corrected 
Higgs masses and CP even mixing parameter $\sin\alpha$ are used to
calculate $\Delta M_s^{LL}/\Delta M_s^{LR}$ from the relation in
eq.~(\ref{suppress}).

\vskip-3mm
\begin{figure}[ht]
\begin{center}
\rotatebox{-90}{\scalebox{0.25}{\includegraphics*{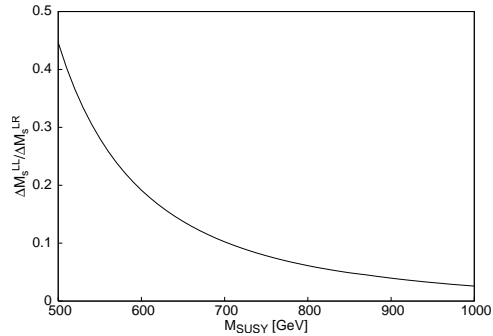}}}
\caption{A plot of the relative suppression of $\Delta M_s^{LL}$ to 
$\Delta M_s^{LR}$ against $M_{SUSY}$. 
The plot is generated using the FeynHiggs package with the
input values, $\tan\beta=50$, $M_{A^0}=500$ GeV, $\mu=1000$ GeV
and $X_t= 1000$ GeV.\label{plot_var_MSusy}}
\end{center}
\end{figure}
The plot in fig.~\ref{plot_var_MA0} shows the ratio of 
$\Delta M_s^{LL}/\Delta M_s^{LR}$ against the pseudoscalar Higgs mass.
For this plot the values $\tan\beta=50$, $\mu=1000$ GeV, $M_{SUSY}=500$ GeV
and $X_t\equiv(A_t-\mu\cot\beta)=1000$ GeV are used.
The size of the suppression is similar to that seen in 
fig.~\ref{plot_DMsLLRL}. For light pseudoscalar Higgs 
mass the ratio is as large as 80\%. 
For a pseudoscalar Higgs mass of 200 GeV the ratio is 45\% and
for a heavy mass the ratio remains at almost 20\%.
The same ratio is shown in fig.~\ref{plot_var_mu} plotted against 
the Higgs mass parameter $\mu$. Here the inputs are the same as 
fig.~\ref{plot_var_MA0} with $M_{A^0}=200$ GeV and $\mu$ allowed to vary.
In this case it is clear that the 
ratio $\Delta M_s^{LL}/\Delta M_s^{LR}$
increases with increasing $\mu$. Fig.~\ref{plot_var_mu} shows that 
for $\mu=1$ TeV  the ratio is 45\% and for $\mu=500$ GeV we still have
a 10\% effect. The plot in fig.~\ref{plot_var_MSusy} shows the variation
of the relative suppression with the SUSY mass scale $M_{SUSY}$.
Again this plot is generated using the same inputs as listed for
fig.~\ref{plot_var_MA0} with $M_{A^0}=200$ GeV and $M_{SUSY}$ allowed
to vary. Here again we see that it is possible for a large contribution
from $\Delta M_s^{LL}$ to exist particularly for light SUSY scales. 
For $M_{SUSY}=500$ GeV there is a 45\% effect which remains at 10\% for
$M_{SUSY}=700$ GeV.

\section{Conclusions}\label{conc}

Using both a simple SUSY SU(5) model and the MSSM Higgs sector with the
FeynHiggs numerical package, 
the Higgs mediated contribution to $\Delta M_s$ in the
large $\tan\beta$ limit has been analysed.
The constraint from the 
recent Tevatron measurement is found to be highly dependant upon the 
determination of $f_{B_s}$. 
It has been quite clearly shown however that there exists large 
regions of the MSSM parameter space for which the contribution from the
operator $(\bar{b}P_L s)(\bar{b}P_L s)$ is non-negligible.
The contribution to this operator may be as large as 80\% of the 
dominant contribution via the operator $(\bar{b}P_L s)(\bar{b}P_R s)$.
Therefore we find that this often ignored operator should 
be considered in any accurate determination of the MSSM 
contribution to the $B_s^0$-$\bar{B}_s^0$ mixing parameter
$\Delta M_s$ in the large $\tan\beta$ regime.



\end{document}